# Effect of magnetism on lattice dynamics in metallic chromium

Stanisław M. Dubiel[1*] and Jan Żukrowski[2]

AGH University of Science and Technology, [1]Faculty of Physics and Applied Computer Science, [2]Academic Center for Materials and Nanotechnology

PL-30-059 Kraków, Poland

## Abstract

Single-crystal sample of chromium doped with ~0.2 at.%[119]Sn was studied by means of a transmission Mössbauer spectroscopy in the temperature range of ~310-315 K. An anomaly in the temperature behavior of the center shift was found at ~313 K, a temperature that coincides well with the Néel temperature of chromium. The anomaly gives evidence that the vibrations of atoms in the studied system are affected by the magnetic state of the sample.

*Corresponding author: Stanislaw.Dubiel@fis.agh.edu.pl (S. M. Dubiel)

Dynamics of lattice in solids has been subject of numerous studies due to its importance both for fundamental and practical aspects in the research of various properties of materials and/or devices. Concerning the former, thermal, elastic and optical properties of solids should be named. Also sound velocity can be explained in terms of the lattice vibrations. Regarding the latter, a noise of electronic devices, can be named as an relevant example. Also different physical quantities like e. g. vibrational entropy, Debye temperature and electron-phonon interactions depend on these vibrations. One of still unresolved problem in the matter of the lattice vibrations is related to the effect of magnetism on these vibrations. It follows from the standard theory of electron-phonon interactions that the effect is insignificant as the ratio between the Debye and the Fermi energy is of the order of $10^{-2}$ [1]. However, from Kim's calculations it follows that for itinerant magnets the electron-phonon coupling can be enhanced by up to two-orders of magnitude [1]. Accordingly, the lattice vibrations in the magnetic phase should be significantly different than the ones in the paramagnetic one. Indeed, inspired by these predictions we have recently carried out Mössbauer spectroscopic measurements on few systems viz. $\sigma$-phase Fe-Cr [2], Fe-V [3] and on C14 $Fe_2Nb$ Laves phase [4] compounds. Magnetism of all these alloy systems is highly itinerant as revealed based on the Rhodes-Wohlfarth plot [5]. Parameters relevant to the lattice dynamics showed significant changes on crossing from a paramagnetic to a ferromagnetic phase. Notably, deviations from the harmonic behavior and/or softening of the lattice were found. It is also worth noticing that recently was reported a giant increase of the average sound velocity in magnetic state of the $\sigma$-FeCr sample [6]. The aim of the present study was to see whether or not the lattice dynamics of metallic chromium is affected by its magnetism which is also greatly itinerant [7]. However, contrary to the previously studied systems [2-5], metallic chromium is an antiferromagnet (AF) with the Néel temperature $T_N \approx 313$ K [8], so it is interesting to examine whether also in this case the lattice vibrations are disturbed by a transition from a paramagnetic into the AF state. To that end a set of $^{119}$Sn-site Mössbauer spectra was recorded in the temperature range of ~310-315 K K on a single-crystal sample of Cr doped by diffusion with ~0.2 at.% $^{119}$Sn, as described elsewhere [9]. Two series of measurements, described in Ref. 10, were performed viz. with an increase and with a decrease of temperature. Examples of the spectra measured in a transmission mode at different temperatures are displayed in Fig. 1.

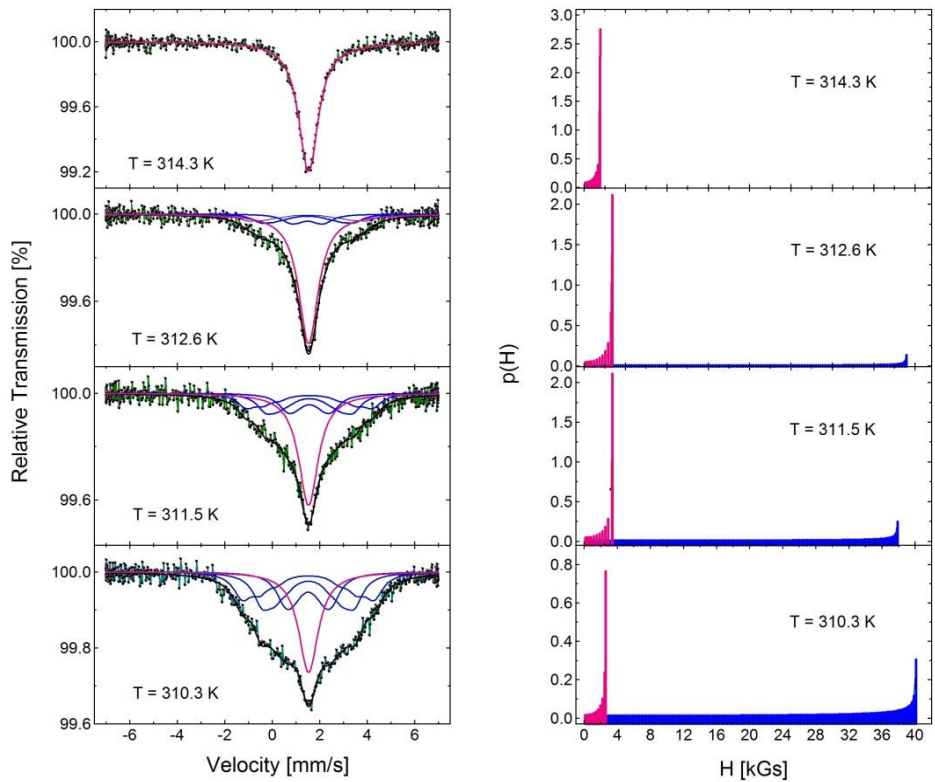

Fig. 1 (left panel) Examples of $^{119}$Sn spectra recorded at various temperatures on the sample of a single-crystal Cr sample doped with ~0.2 at.% $^{119}$Sn. Subspectra involved in their analysis with method A are indicated; (right panel) Histograms of the hyperfine field distribution as obtained with the fitting procedure B.

The spectra were analyzed using two different procedures A and B. In procedure A it was assumed that each spectrum is composed of a magnetic and a paramagnetic component. The former could have been well fitted in terms of three sextets and the latter by a single line. Each sextet was characterized by the hyperfine field, $H_k$ (k=1,2,3), center shift, CS (common to all), line width, G (common to all), relative abundance, $A_k$(k=1,2,3). Relative ratio of lines within each sextet was fixed to 3:2:1. Spectral parameters of the singlet were: center shift, CS, line width, G and relative abundance, A. In the procedure B the spectra were analyzed in terms of two fundamental spin-density waves (SDWs) having different amplitudes (spin-density or hyperfine fields) and the same charge-densities (isomer shift or center shift). This approach follows from the fact that (1) AF of chromium is constituted by SDWs and the fundamental harmonic amounts to ~99% around $T_N$, and (2) in all the spectra recorded in the vicinity of $T_N$ for $T < T_N$ a single-line contribution can be seen. This likely means that there are two magnetic domains with different value of $T_N$ present in the investigated sample. It is well known that strain in Cr affects $T_N$. Thus existence of different strain distribution across the sample would result in different values of T. Both fitting procedures resulted in good fits i.e. values of the chi-squared and those of the misfit were statistically equivalent. However, from the point of view of the underlying physics, the procedure B is, in our opinion, more justified. The parameter of merit relevant to the lattice vibration is a temperature dependence of the center shift, CS(T), which according the Debye model is given by the following equation:

$$CS(T) = IS(0) - \frac{3k_B T}{2mc}\left[\frac{3T_D}{8T} + \left(\frac{T}{T_D}\right)^3 \int_0^{T_D/T} \frac{x^3}{e^x - 1} dx\right] \qquad (1)$$

Where *IS(0)* stays for the isomer shift (temperature independent), $k_B$ is the Boltzmann constant, *m* is a mass of $^{119}$Sn atoms, $T_D$ is the Debye temperature and the second term is known as the second-order Doppler term, $\delta_{SOD}(T)$.

The CS(T) dependence obtained from the fitting procedure A is shown in Fig. 2 while the one retrieved based on the procedure B can be seen in Fig. 3.

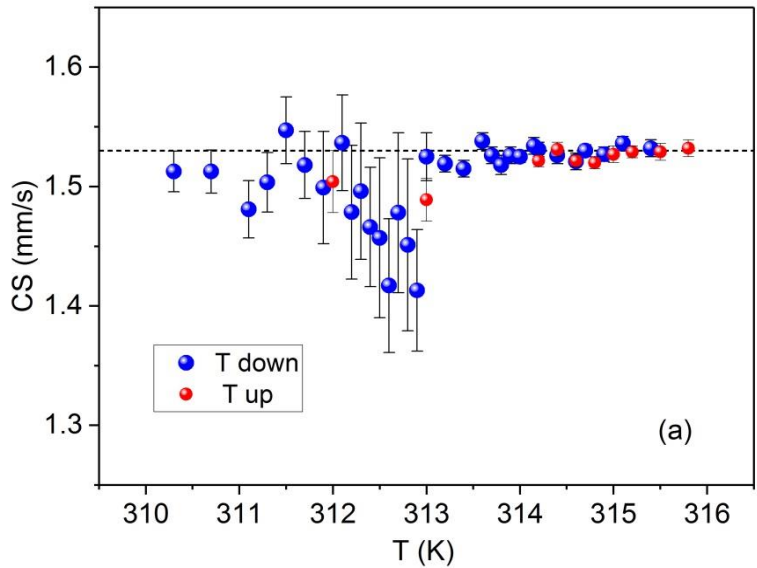

Fig. 2 Temperature dependence of the center shift, CS, as found with the procedure A. The horizontal line stays for the CS-value in the paramagnetic state.

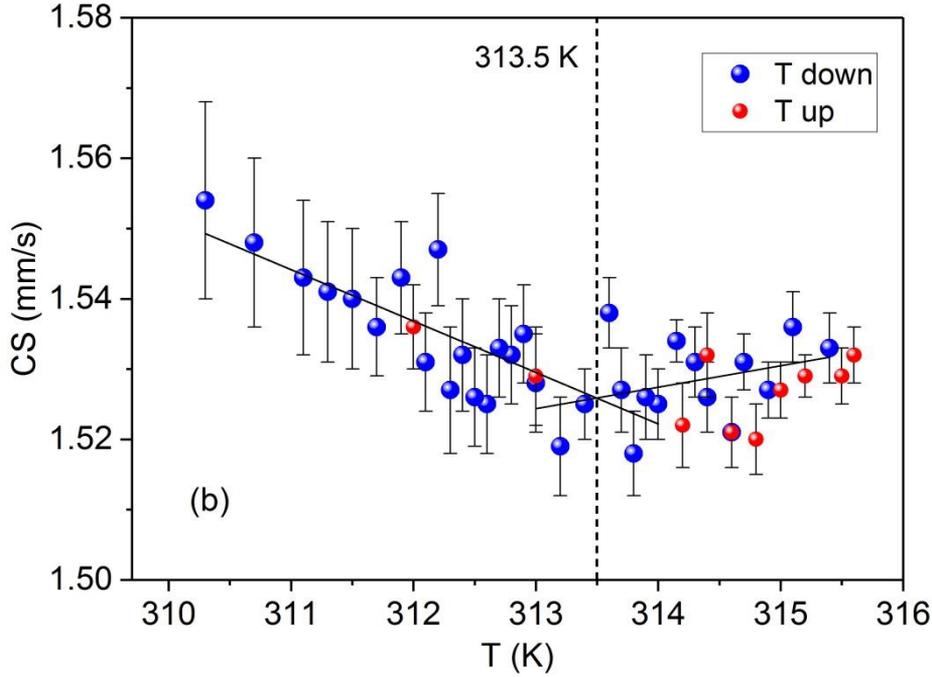

Fig. 3 Temperature dependence of the center shift, CS, as found with the procedure B. Solid lines stay for the best linear fit to the data in the paramagnetic and in the antiferromagnetic state, respectively. The vertical dashed line indicates the crossing of the two straight lines.

Both Fig. 2 and Fig. 3 give evidence that there is an anomaly at T≈313 K which coincides well with the Néel temperature.

As the fitting procedure B is from the underlying physics more justified and the anomaly revealed based on this procedure is statistically better evidenced, we will discuss it in a more quantitative way. As is well known, the temperature change of CS has its origin in the so-called second-order Doppler shift, $\delta_{SOD}$, that is related to the average of the square velocity of vibrations, $<v^2>$, by the following equation:

$$\delta_{SOD}(T) = -<v^2>(T)/2c \qquad (2)$$

where $c$ stays for the velocity of light. Thus a change of $\delta_{SOD}$, $\Delta\delta_{SOD}$, caused by a change of temperature, $\Delta T$, corresponds to the underlying change of the average squared velocity, $\Delta <v^2>$:

$$\Delta <v^2> = -2c \cdot \Delta\delta_{SOD} \qquad (3)$$

Having determined $\Delta <v^2>$ one can figure out the related change of the kinetic energy of vibrations, $\Delta E_k$:

$$\Delta E_k = 0.5 \cdot m \cdot \Delta <v^2> \qquad (4)$$

In this way we have found that the change of CS(T) in the temperature range 310.3 – 313.5 K (antiferromagnetic state) corresponds to a decrease of the kinetic energy of vibrations by 8.5 meV, whereas the change of CS(T) in the temperature interval of 313.5-315.6 K (paramagnetic state) is equivalent to an increase of the kinetic energy of vibrations by 2.2 meV. In other words, the kinetic energy of vibrations in the paramagnetic state close to the Néel temperature on lowering T increases at the rate of ~1meV/K while this energy on lowering T decreases in the antiferromagnetic state close to the Néel temperature at the rate of ~2.7 meV/K. The corresponding changes of energy due to the decrease of temperature, $\Delta E = k_B \cdot \Delta T$, are -0.27 meV and -0.18 meV, respectively.

These results give clear evidence that the lattice vibrations in metallic chromium are significantly affected by the magnetism of the system and the effect goes in opposite directions above and below $T_N$. One may doubt that the present experiment does not give a direct evidence on the lattice vibrations of Cr atoms themselves because the measurements were performed on Sn atoms introduced into the Cr matrix. So consequently, the vibrations of Sn atoms, which are much heavier than the matrix Cr atoms, may be different. We want to make the following remarks to this issue. Based on our previous studies of magnetism of chromium using the [119]Sn Mössbauer spectroscopy we found that (1) the third-order harmonics of the SDWs has amplitude and sign [9] in accord with the corresponding values revealed by the neutron (ND) diffraction study, (2) the value of the Néel temperature agrees within ±1 K [10] with the value found from the ND study, (3) the value of the spin-flip temperature [11] agrees well with that determined with ND, and (4) the anomaly in CS(T) revealed in the present study occurs at the temperature that coincides with $T_N$ of a pure Cr alloy.

Based on all these findings we can quite confidently conclude that the vibrations of Sn atoms embedded into the Cr matrix correctly reproduce the underlying vibrations of the Cr matrix. In other words, the anomaly in vibrations of the probe Sn atoms detected in the present experiment unambiguously reflects the anomaly in the vibrations of the matrix atoms observed on a transition from a paramagnetic into the antiferromagnetic state of the system.

**CRediT authorship contribution statement**

**Acknowledgements**

This work was financed by the Faculty of Physics and Applied Computer Science AGH UST and ACMIN AGH UST statutory tasks within subsidy of Ministry of Science and Higher Education, Warszawa.

**References**

[1] D. J. Kim, Phys. Rev. B, 25 (1982) 6919; Phys. Rep., **171** (1988) 129

[2] S. M. Dubiel, J. Cieślak, M. Reissner, EPL, **101** (2013) 16008

[3] S. M. Dubiel and J. Żukrowski, J. Magn. Magn. Mater., **441** (2017) 557

[4] J. Żukrowski and S. M. Dubiel, J. Appl. Phys., **123** (2018) 223902

[5] P. R. Rhodes and E. P. Wohlfarth: Proc. R. Soc. London A **273** (1963) 247

[6] S. M. Dubiel and A. I. Chumakov, EPL, **117** (2017) 56001

[7] E. Fawcett, Rev. Mod. Phys., **60** (1988) 209.

[8] R. D. Metcalfe, G. A. Saunders, M. Cankurtaran et al., Philos. Mag. B, **79** (1999) 663.

[9] S. M. Dubiel and G. LeCaër G., Europhys. Lett., **4** (1987) 487

[10] S. M. Dubiel and J. Cieślak, Europhys. Lett., **53** (2001) 383

[11] S. M. Dubiel, Rev. Phys. B, **29** (1984) 2818